# IS WAVE MECHANICS CONSISTENT WITH CLASSICAL LOGIC?

**Adriano Orefice**[*]**, Raffaele Giovanelli, Domenico Ditto**

*University of Milan – DISAA. Via Celoria, 2 - 20133 - Milano (Italy)*

ABSTRACT - Contrary to a wide-spread commonplace, an exact, ray-based treatment holding for any kind of monochromatic wave-like features (such as diffraction and interference) is provided by the structure itself of the Helmholtz equation. This observation allows to dispel - in *apparent* violation of the Uncertainty Principle - another commonplace, forbidding an exact, trajectory-based approach to Wave Mechanics.

## 1 – Introduction

Mainly because of the statistical Copenhagen interpretation given to de Broglie's matter waves, and because of the Uncertainty Principle, Wave Mechanics appeared to require the abandonment of the classical idea of particle trajectories *and of the logic itself of Classical Mechanics.*
The recent demonstration, however, that an exact, ray-based description is possible for any kind of monochromatic features described by the *Helmholtz* equation [**1-4**] opens the way to an exact, trajectory-based approach to Wave Mechanics based on the *Helmholtz-like* time-independent Schrödinger equation and running as close as possible to Classical Mechanics.
Given that the Uncertainty Principle *is not a necessary pre-requisite of Wave Mechanics,* we are interested here in finding out what would Wave Mechanics "say" *if the logic of Classical Mechanics were restored.*
We consider in Sects.2 and 3 the strictly correlated Hamiltonian trajectory systems holding, respectively, in classical and wave-mechanical cases, and *apparently* violating, in the latter case, the Uncertainty Principle. We discuss in Sects.4 and 5 their connection with the Copenhagen and Bohmian interpretations, and compare in Sect.6 the cases of exact and statistical "quantum trajectories".

## 2- Exact Hamiltonian ray kinematics of classical Helmholtz waves

We shall assume, in the following, both *wave monochromaticity* and *stationary media*, and refer in the present Section, in order to fix ideas, to *classical electromagnetic waves* described by a scalar wave equation of the simple form

---

[*] Corresponding author - adriano.orefice@unimi.it

$$\nabla^2 \psi - \frac{n^2}{c^2} \frac{\partial^2 \psi}{\partial t^2} = 0 \ , \tag{1}$$

where $\psi(\vec{r},t)$ represents any component of the electric and/or magnetic field, and $n$ is the (time independent) refractive index of the medium. By assuming

$$\psi = u(\vec{r},\omega) e^{-i\omega t} \ , \tag{2}$$

we get from eq.(1) the Helmholtz equation [5]

$$\nabla^2 u + (n k_0)^2 u = 0 \tag{3}$$

(with $k_0 \equiv \frac{2\pi}{\lambda_0} = \frac{\omega}{c}$), and look for solutions of the (quite general) form

$$u(\vec{r},\omega) = R(\vec{r},\omega) \ e^{i \varphi(\vec{r},\omega)} \ , \tag{4}$$

where the real functions $R(\vec{r},\omega)$ and $\varphi(\vec{r},\omega)$ represent respectively, *without any probabilistic meaning*, the amplitude and phase of the monochromatic waves to be dealt with. No use is made of *plane* monochromatic waves: the (time-independent) phase surfaces $\varphi(\vec{r},\omega) = const$ are determined, as we shall see, by the boundary conditions. After the introduction of a *wave vector*

$$\vec{k} = \vec{\nabla} \varphi(\vec{r},\omega) \ , \tag{5}$$

of a *"Wave Potential"* function

$$W(\vec{r},\omega) = -\frac{c}{2 k_0} \frac{\nabla^2 R(\vec{r},\omega)}{R(\vec{r},\omega)} \tag{6}$$

and of a "structural" function

$$D(\vec{r},\vec{k},\omega) \equiv \frac{c}{2 k_0} [k^2 - (n k_0)^2] + W(\vec{r},\omega) \tag{7}$$

describing the physical frame of the problem, eq.(3) turns out to be associated [1-4] with the *closed* and *exact* Hamiltonian system

$$\begin{cases} \dfrac{d\vec{r}}{dt} = \dfrac{\partial D}{\partial \vec{k}} \equiv \dfrac{c\vec{k}}{k_0} & (8) \\[6pt] \dfrac{d\vec{k}}{dt} = -\dfrac{\partial D}{\partial \vec{r}} \equiv \vec{\nabla} [\dfrac{c k_0}{2} n^2(\vec{r},\omega) - W(\vec{r},\omega)] & (9) \\[6pt] \vec{\nabla} \cdot (R^2 \vec{k}) = 0 & (10) \end{cases}$$

providing the kinematics (i.e. both geometry and motion laws) of monochromatic rays moving along a *stationary* set of trajectories with the "intrinsic" ray velocity $\frac{d\vec{r}}{dt} = \frac{c\vec{k}}{k_0}$. These "*Helmholtz ray trajectories*" are mutually coupled by the Wave Potential function $W(\vec{r},\omega)$ ("triggered" by the wave amplitude distribution $R(\vec{r},\omega)$ over the phase fronts) which is encoded in the Helmholtz equation itself, and represents *the one and only cause of wave-like features such as diffraction and interference*.

Once assigned the starting values $\vec{r}(\omega, t=0)$ and $\vec{k}(\omega, t=0)$, respectively, of the ray positions and wave vectors, together with the wave amplitude distribution $R(\vec{r}, \omega)$ over a suitable "launching" surface (Fig.1), the double role of eq.(10), in the numerical integration of the Hamiltonian system, is

- to provide, step by step, the *necessary and sufficient condition* for the determination of $\vec{r}(t)$, $\vec{k}(t)$ and $R(\vec{r},\omega)$ over the next wave-front (Fig.1), thus allowing a self-consistent "closure" of the kinematic system, and
- to show that the coupling term $\vec{\nabla} W(\vec{r},\omega)$ acts, at each point, *perpendicularly* to the relevant (monochromatic) ray trajectories, thus *confining its action to a mere deflection*.

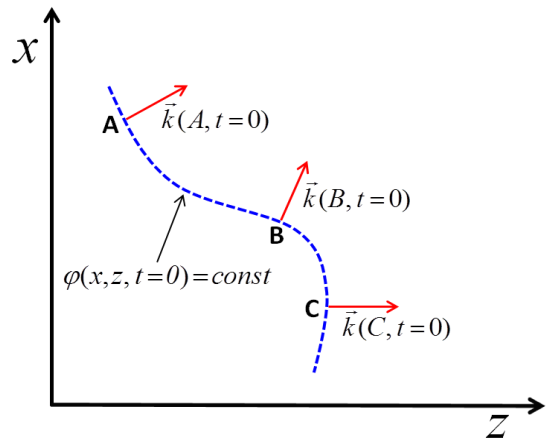

Fig.1 Wave launching stage

The limit of *geometrical optics* is allowed when the rays are no longer appreciably coupled by Wave Potential, and propagate independently from one another under the only influence of the refractive index of the medium, without being subject to any kind of wave-like phenomena such as diffraction and/or interference.

In conclusion, contrary to the commonplace that a treatment in terms of *ray-trajectories* is only possible for a limited number of physical cases (such as reflection and refraction) ascribed to the *geometrical optics approximation,* eqs.(8)-(10) provide an exact, ray-based description of any kind of wave-like features. We present in Figs.2 and 3 the numerical application to rays starting from $z = 0$, parallel to the $z$-axis, and undergoing *diffraction* through slits of half width $w_0$, with $\varepsilon \equiv \lambda_0 / w_0 < 1$. The problem is faced *by taking* into account, for simplicity sake, the ray trajectories of electromagnetic waves *in vacuo* $(n^2 = 1)$, with a geometry allowing to limit the computation to the $(x, z)$-plane and *by* *solving*, step by step, the Hamiltonian system (8)-(10) by means of a symplectic numerical integration method.

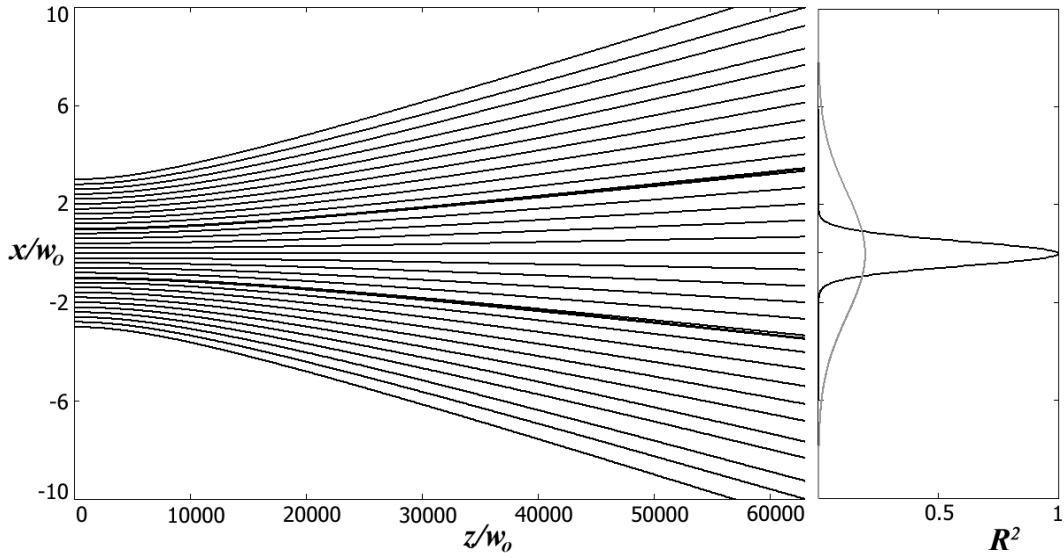

Fig.2 - Gaussian beam diffraction

Fig.2 refers to the *diffraction* of a *Gaussian* beam launched along *z* from a vertical slit centered at $x = z = 0$ in the form $R(x; z = 0) \div exp(-x^2/w_0^2)$. We plot on the *left-hand* side the ray-trajectory patterns, and on the *right-hand* side the initial and final transverse intensity distributions of the wave. The two heavy lines represent the analytical *paraxial* approximation [6]

$$x(z) = \pm \sqrt{w_0^2 + \left(\frac{\lambda_0 \, z}{\pi \, w_0}\right)^2} \qquad (11)$$

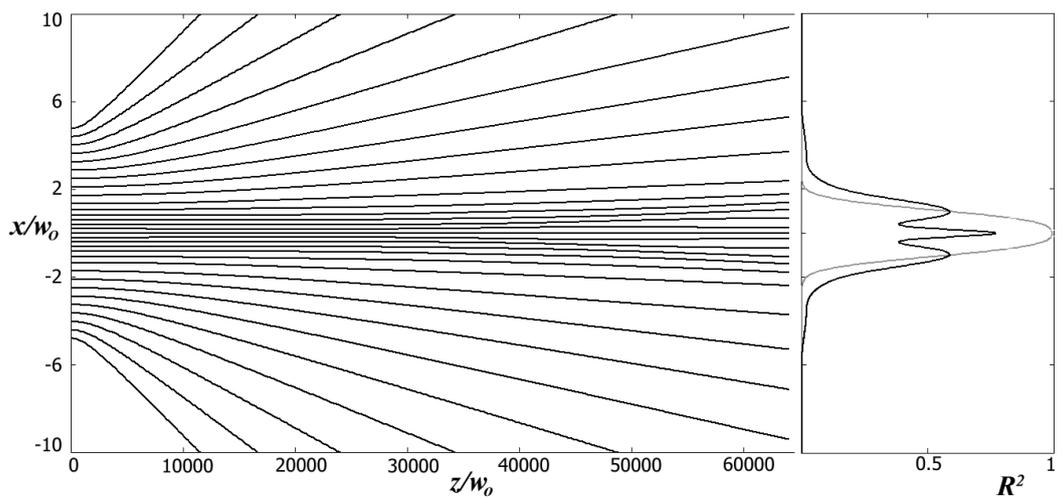

Fig.3 - Non-Gaussian beam diffraction

of the trajectories starting (at $z = 0$) from the so-called "*waist*" *positions* $x/w_0 = \pm 1$ . The excellent agreement between the *analytical* eq.(11) and our *numerical* results provides, of course, a test of our approach and interpretation.
Fig.3 represents, in its turn, the trajectory pattern (and the initial and final transverse intensity distributions) in the case of *non-Gaussian beam diffraction*.
It's worthwhile recalling that while the equation system (8-10) provides an *exact* Hamiltonian description of the wave kinematics, an *approximate* ray-tracing (based on a *complex eikonal equation*, amounting to a first-order approximation of the wave beam diffraction) was presented in 1993/94 by one of the Authors (A.O., [**7**,**8**]), for the *quasi-optical* propagation of electromagnetic *Gaussian beams* at the electron-cyclotron frequency in the magnetized plasmas of Tokamaks such as JET and FTU, and applied in recent years [**9**] by an *équipe* working on the Doppler back-scattering microwave diagnostics installed on the Tokamak TORE SUPRA of Cadarache.

### 3- Exact Hamiltonian particle dynamics in Wave Mechanics

Let us pass now to the case of non-interacting, spinless particles of mass *m* and assigned total energy *E*, launched with an initial momentum $\vec{p}_0$ (with $p_0 = \sqrt{2mE}$ ) into a force field deriving from a *stationary* potential energy $V(\vec{r})$.
The *classical* dynamical behavior of each particle is described, as is well known, by the time-independent Hamilton-Jacobi equation

$$(\vec{\nabla} S)^2 = 2m[E - V(\vec{r})] \ , \qquad (12)$$

where the basic property of the function $S(\vec{r}, E)$ is that the particle momentum is given by

$$\vec{p} = \vec{\nabla} S(\vec{r}, E). \qquad (13)$$

In other words, the (time-independent) Hamilton-Jacobi surfaces $S(\vec{r}, E) = const$ are perpendicular to the momentum of the moving particles, and *pilot* them along *stationary* trajectories according to their dynamical motion laws.
Considerations based on the variational principles both of Maupertuis and Fermat [**5**] induced Louis de Broglie [**10**, **11**] to associate material particles with suitable "*matter waves*", according to the correspondence

$$\vec{k} \equiv \vec{\nabla} \varphi \leftrightarrow \vec{p}/\hbar \equiv \vec{\nabla} S(\vec{r}, E)/\hbar \qquad (14)$$

*viewing the Hamilton-Jacobi surfaces* $S(\vec{r}, E) = const$ *as the phase-fronts of mono-energetic matter waves, while maintaining the piloting role played in Classical Mechanics.*
The successive step, due to Schrödinger [**12**-**15**], may be simply performed [**4**] by viewing Classical Mechanics - represented here by eq.(12) - as the geometrical optics approximation of a Helmholtz-like equation holding for de Broglie's matter waves, thus obtaining the *time-independent equation*

$$\nabla^2 u(\vec{r},E) + \frac{2m}{\hbar^2}[E - V(\vec{r})]\, u(\vec{r},E) = 0, \tag{15}$$

holding for matter waves associated with particles of total energy *E* moving in a *stationary* potential field $V(\vec{r})$.

As is well known, the *objective physical existence* of de Broglie's waves was very soon revealed by the Davisson and Germer experiments of electron diffraction in crystalline nickel targets [**16**].

The same mathematical procedure applied in the previous (classical) Section to the Helmholtz eq.(3) may now be applied to the Helmholtz-like eq.(15), in order to search for a stationary set of *exact particle trajectories* corresponding to the ray trajectories of the previous Section. Recalling eqs. (4), (5) and (14) we look therefore for solutions of eq.(15) of the form

$$u(\vec{r},E) = R(\vec{r},E)\, e^{i\, S(\vec{r},E)/\hbar}, \tag{16}$$

where the real functions $R(\vec{r},E)$ and $S(\vec{r},E)$ represent (without any probabilistic meaning) the amplitude and phase of matter waves.

After separation of real and imaginary parts, and after having introduced the function

$$Q(\vec{r},E) = -\frac{\hbar^2}{2m}\frac{\nabla^2 R(\vec{r},E)}{R(\vec{r},E)}, \tag{17}$$

strictly analogous to the Wave Potential of eq.(6), one obtains [**1-4**] the "structural" Hamiltonian relation

$$H(\vec{r},\vec{p},E) \equiv \frac{p^2}{2m} + V(\vec{r}) + Q(\vec{r},E) = E \tag{18}$$

whose differentiation $\frac{\partial H}{\partial \vec{r}} \cdot d\vec{r} + \frac{\partial H}{\partial \vec{p}} \cdot d\vec{p} = 0$ shows that the time-independent Schrödinger equation (15) is associated with the *exact* and *self-consistent* Hamiltonian dynamical system, strictly analogous to the system (8)-(10),

$$\begin{cases} \dfrac{d\vec{r}}{dt} = \dfrac{\partial H}{\partial \vec{p}} \equiv \dfrac{\vec{p}}{m} & (19) \\[4pt] \dfrac{d\vec{p}}{dt} = -\dfrac{\partial H}{\partial \vec{r}} \equiv -\vec{\nabla}[V(\vec{r}) + Q(\vec{r},E)] & (20) \\[4pt] \vec{\nabla} \cdot (R^2\, \vec{p}) = 0 & (21) \end{cases}$$

Once assigned the starting values $\vec{r}(E,t=0)$ and $\vec{p}(E,t=0)$ of the particle position and momentum, together with the wave amplitude distribution $R(\vec{r},E)$ over a suitable launching surface, the time-integration of the system provides the values of $\vec{r}(E,t)$ and $\vec{p}(E,t)$, i.e. a full description of the particle dynamics along a stationary set of "*Helmholtz trajectories*".

The simultaneous assignment of $\vec{r}(E,t=0)$ and $\vec{p}(E,t=0)$ is performed in *apparent* violation of the Uncertainty Principle.
This is done both because of the strict mathematical analogy of the present case with the classical case of the previous Section, and because of the fact that such a Principle *is not a necessary pre-requisite of Wave Mechanics*. As was bravely observed, in fact, in Ref.[17], with reference to the well-known relations

$$\begin{cases} \Delta x\, \Delta p_x \geq \hbar \\ \Delta y\, \Delta p_y \geq \hbar \\ \Delta z\, \Delta p_z \geq \hbar \end{cases},$$

holding, in agreement with the Uncertainty Principle, for the particle description in terms of wave-packets, "*the Uncertainty Principle is not a result particular of quantum mechanics, but general of any wave theory, since $\Delta x, \Delta y, \Delta z$ and $\Delta p_x, \Delta p_y, \Delta p_z$ only measure the dispersion associated with the wave, without specifying the physics described by such a wave*".
*At exactly the same logical level*, when particles are described by Helmholtz-like equations such as the Schrödinger and Klein-Gordon time-independent equations [**1-4**], exact *Helmholtz trajectories* represent an alternative choice "*general of any wave theory*". We are interested therefore in these trajectories, i.e. in finding out what would Wave Mechanics say if the logic of Classical Mechanics were preserved without resorting to further restrictions.
The function $Q(\vec{r},E))$ - which we call once more, for simplicity sake, "*Wave Potential*" - has the same basic structure and coupling role of the function $W(\vec{r},\omega) = -\dfrac{c}{2k_0}\dfrac{\nabla^2 R(\vec{r},\omega)}{R(\vec{r},\omega)}$ of the previous Section, so that its *presence* is, once more, the one and only cause of diffraction and/or interference**,** while its *absence* would reduce the system (19)-(21) to the standard set of *classical dynamical* equations, which constitute therefore, as expected, its *geometrical optics* approximation. The coupled particle motion turns out to occur in a way somewhat reminding a large railway system, where a *fixed* pattern of tracks is laid according to an overall structural design and is run by "wagons" according to an assigned time-table. Diffraction and interference patterns are a part of this stationary structure, along which the "wagons" move in a merely geometrical "entanglement".
In complete analogy with the case of eq.(10) of the previous Section, eq.(21) has the double role of

- providing both $R(\vec{r},E)$ and $Q(\vec{r},E)$ along the full set of coupled trajectories, thus allowing the self-consistent "closure" of the wave-dynamical Hamiltonian system, and
- showing that the "force" term $\vec{\nabla} Q(\vec{r},E)$ maintains itself *perpendicular* to the particle trajectories, so that *no energy exchange* is involved in its merely deflecting action.

As far as numerical applications are concerned, it may be observed that the Hamiltonian systems (8)-(10) and (19)-(21) turn out to be formally coincident [**1-4**] when written in suitable adimensional form, leading therefore to *the same numerical solutions*, which we shall not repeat.

It's also worthwhile to remind that while eqs.(19)-(21) provide an *exact* and general Hamiltonian description of the particle motion, an *approximate* treatment was presented by one of the Authors (A.O., [**18**]), for the particular case of *Gaussian* electron beams. A *complex eikonal* equation, amounting to a first order quasi-optical approximation of the *quantum* particle diffraction, was adopted there - in complete analogy with the *classical* electromagnetic case of Refs.[**7-9**] - in order to overcome the collapse, for narrow beams, of the ordinary, zero-order, *real eikonal* approximation.

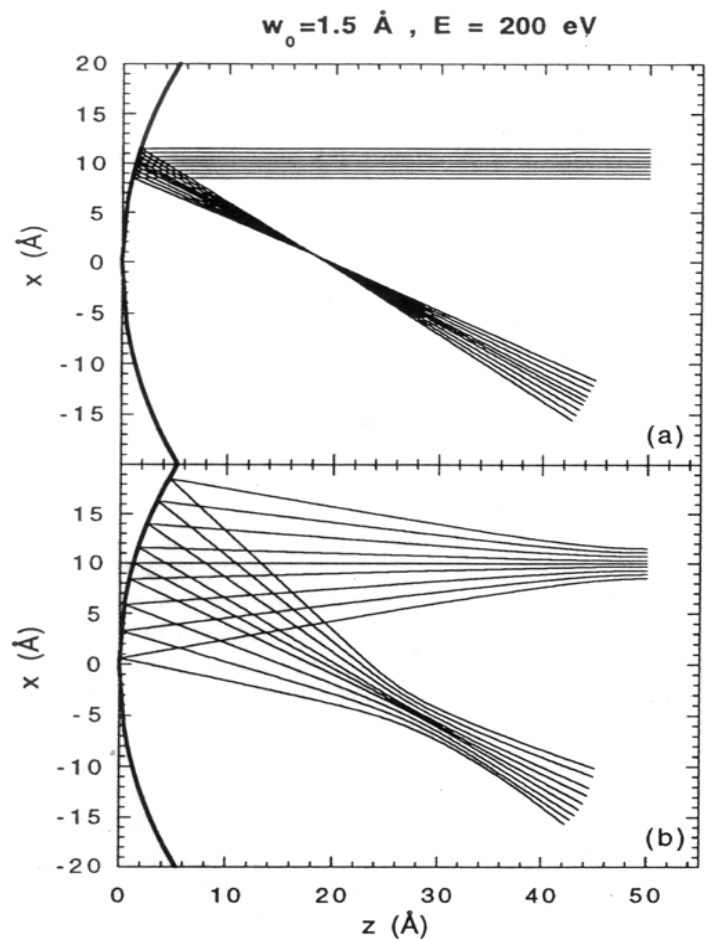

Fig.4 - Gaussian electron beam launched against an electrostatic mirror

We present in Fig.4, borrowed from Ref.[**18**], the case of a collimated *Gaussian* electron beam launched from the right-hand side into an assigned electrostatic potential acting as a spherical mirror, both (**a**) *neglecting* and (**b**) *taking into account* the Wave Potential term due to its Gaussian amplitude distribution. In case (**a**) the beam of de Broglie's matter waves is reflected and focused by the electrostatic mirror according to the *geometrical optics approximation*, while in case (**b**) the focus is replaced by a finite "waist".

## 4 - The Copenhagen interpretation

Let us now recall that, starting from eqs.(2) and (15) and assuming the "quantum" relation

$$E = \hbar\omega, \qquad (22)$$

one obtains [**14**] the ordinary-looking wave equation

$$\nabla^2\psi = \frac{2m}{E^2}[E-V(\vec{r})]\frac{\partial^2\psi}{\partial t^2}, \qquad (23)$$

describing the propagation and dispersive character of *mono-energetic* de Broglie matter waves. By means, moreover, of the same eqs. (2), (15) and (22) one may also get [**12-15**] the equation

$$\nabla^2\psi - \frac{2m}{\hbar^2}V(\vec{r})\psi = -\frac{2m}{\hbar^2}E\psi \equiv -\frac{2mi}{\hbar}\frac{E}{\hbar\omega}\frac{\partial\psi}{\partial t} = -\frac{2mi}{\hbar}\frac{\partial\psi}{\partial t}, \qquad (24)$$

which is the usual form of the *time-dependent Schrödinger equation* for particles moving in a stationary potential field $V(\vec{r})$. Any wave-like implication of eq.(24) is due to its connection with the *time-independent* Schrödinger equation (15), *from which it is obtained*. Eq.(15) admits indeed, as is well known, a (discrete or continuous, according to the boundary conditions) set of energy eigen-values and ortho-normal eigen-modes, which (referring for simplicity to the discrete case) we indicate respectively by $E_n$ and $u_n(\vec{r})$; and, making use of eqs.(2) and (22) and defining both the *eigen-frequencies* $\omega_n \equiv E_n/\hbar$ and the *eigen-functions*

$$\psi_n(\vec{r},t) = u_n(\vec{r})e^{-i\omega_n t} \equiv u_n(\vec{r})e^{-iE_n t/\hbar}, \qquad (25)$$

it's a standard procedure to verify that any linear superposition (with arbitrary constant coefficients $c_n$) of the form

$$\psi(\vec{r},t) = \sum_n c_n \psi_n(\vec{r},t), \qquad (26)$$

is a solution of eq.(24), describing therefore, in general, the deterministic evolution of an *arbitrary superposition* of mono-energetic eigen-waves $\psi_n(\vec{r},t)$, each one of which travels according to a wave equation of the form (23) (with $E = E_n$) along the *Helmholtz trajectories* associated with the relevant *time-independent* Schrödinger eq.(15).

Max Born proposed for the function (26), as is well known, a role [**19**] going much beyond that of a simple superposition: *although eq.(24) is not - in itself - a wave equation*, its solution (26) was called "Wave-Function", and assumed to represent *the most complete possible description of the physical state of a particle whose energy is not determined*. Even though "no generally accepted derivation has been given to date" [**20**], this "Born Rule" aroused, together with Heisenberg's uncertainty relations, an almost universally accepted *probabilistic* conception of physical reality, associating moreover to the continuous and deterministic evolution of $\psi(\vec{r},t)$ provided by eq.(24) the further postulate of a discontinuous and probability-dominated evolution process, after interaction with a measuring apparatus, in the form of a *collapse* (according to the probabilities $|c_n|^2$, in duly normalized form) into a single eigen-state.

In default of the mathematical tools (drawn from the standard time-independent Schrödinger equation) exploited in our present approach, the behavior of a particle is usually described, in the Copenhagen interpretation, by a packet of plane wave trains. A particle of undetermined energy is viewed, for *V=0*, as associated with a narrow wave-packet [**14**] of the form

$$\psi(\vec{r},t) = \int c(\vec{p})\, e^{\frac{i}{\hbar}[\vec{p}\cdot\vec{r} - (\frac{p^2}{2m})t]} d\vec{p} \qquad (27)$$

displacing itself (while deforming) with a *group velocity*

$$\vec{v}_g \equiv \frac{d\omega}{d\vec{k}} = \frac{d(E/\hbar)}{d(\vec{p}/\hbar)} = \frac{d(p^2/2m)}{d\vec{p}} = \frac{\vec{p}}{m} \qquad (28)$$

whose expression suggests its identification with the *particle velocity* of Classical Mechanics. In Born's words [**21**] "*this identification is very attractive: in particular it tempts us to interpret a particle of matter as a wave-packet due to the superposition of a number of wave trains. But this tentative interpretation comes up against insurmountable difficulties, since a wave packet of this kind is in general very soon dissipated*". The "dissipation" is due, of course, to the dispersive character of eq.(23).

A wave-packet doesn't represent however, as we have shown, the only available option: the particle may be consistently associated, in fact, with a *monochromatic wave*, allowing to describe its motion in terms of exact *Helmholtz trajectories* along which it moves with a velocity which is given - rather than by the ambiguous eq.(28) - by the classical-looking eq.(19) of the present paper: $\frac{d\vec{r}}{dt} = \frac{\partial H}{\partial \vec{p}} \equiv \frac{\vec{p}}{m}$, *without any dispersion and uncertainty.*

**5** - **The Bohmian approach**

Let us finally come to the case of Bohm's theory **[22-33],** and to its connection with the present analysis. In Bohm's approach [**22**] a replacement of the form

$$\psi(\vec{r},t) = R(\vec{r},t)\, e^{i G(\vec{r},t)/\hbar}, \qquad (30)$$

with real $R(\vec{r},t)$ and $G(\vec{r},t)$, leads, when introduced into eq.(24), to a system of coupled fluid-like equations (which we shall omit here for brevity sake) whose coupling role is played by a time-dependent "Quantum Potential" term of the form

$$Q_B(\vec{r},t) = -\frac{\hbar^2}{2m}\frac{\nabla^2 R(\vec{r},t)}{R(\vec{r},t)}, \qquad (31)$$

formally coinciding with the mono-energetic, time-independent "Wave Potential" $Q(\vec{r},E) = -\frac{\hbar^2}{2m}\frac{\nabla^2 R(\vec{r},E)}{R(\vec{r},E)}$ of eq.(17), of which it represents in fact a time-evolving average over the whole set of eigen-functions (26).

Bohm's replacement (30) - shaped on eq.(16), i.e on de Broglie's *mono-energetic* waves, whose objective reality was established once and for all by the Davisson and Germer experiments - *aims to dress with plausibility the Born Rule* by depicting $\psi(\vec{r},t)$ as a *single wave* (which it surely is not) hopefully sharing the same experimental evidence of de Broglie's waves, and even as the most general "pilot-wave". According, indeed, to Ref.[**29**], "*Born had an absolutely correct (...) intuition about the meaning of the wave function, namely that it guides the particles and it determines the probability of particle positions (...). Born is close to Bohmian mechanics*".

The explicit use of the Quantum Potential term is formally bypassed in modern Bohmian Mechanics by the assumption of the "guidance equation"

$$\frac{d\vec{r}(t)}{dt} = \vec{\nabla}G(\vec{r},t)/m \equiv \frac{\hbar}{mi}\,Im\,(\frac{\vec{\nabla}\psi}{\psi}) \equiv \frac{\hbar}{2mi}\frac{\psi^*\vec{\nabla}\psi - \psi\vec{\nabla}\psi^*}{\psi\psi^*}, \qquad (32)$$

where $\psi\psi^* \equiv |\psi|^2 \equiv R^2$, and the expression of $\vec{\nabla}G(\vec{r},t)$ is immediately obtained from eq.(30). The time-integration of eq.(32) is made possible by the feedback input, step by step, of the value assumed by $\psi(\vec{r},t)$ in the simultaneous solution of the relevant *time-dependent* Schrödinger equation; and the results are *apodictically* claimed to represent the "quantum trajectories" of each particle. Since however, as is shown in any textbook of Quantum Mechanics [**14, 15**], the quantity $\vec{J} \equiv \frac{\hbar}{2mi}(\psi^*\vec{\nabla}\psi - \psi\vec{\nabla}\psi^*)$ represents a *probability current density* (whose statistical significance is shared even by stationary states) the "guidance velocity" $\frac{d\vec{r}(t)}{dt}$ turns out to coincide with $\vec{J}/R^2$, limiting itself to represent "*the flux lines along which the probability density is transported*" [**29**]. In Bohm's words, indeed, the quantity $R^2$ is assumed to represent - in the attempt to deviate as little as possible from the Copenhagen interpretation - "*the probability density for particles belonging to a statistical ensemble*" [**22**], thus laying an unavoidably statistical imprint on the whole of his theory. It is quite symptomatic that no objection was raised about the consistency of Bohmian "quantum trajectories" with the Uncertainty Principle.

**6 - Discussion and conclusions**

We summarize our comparison in Tables I and II, holding for particles moving in a stationary potential $V(\vec{r})$ .

| TAB.I<br><br>EXACT (POINT-PARTICLE)<br>DESCRIPTION | TAB.II<br><br>BOHMIAN (WAVE-PACKET)<br>DESCRIPTION |
|---|---|
| $$\frac{d\vec{r}}{dt} = \frac{\vec{p}}{m}$$ $$\frac{d\vec{p}}{dt} = -\vec{\nabla}[\,V(\vec{r}) - \frac{\hbar^2}{2m}\frac{\nabla^2 R(\vec{r},E)}{R(\vec{r},E)}\,]$$ $$\vec{\nabla}\cdot(R^2\,\vec{p}) = 0$$ | $$\frac{d\vec{r}}{dt} = \frac{\hbar}{mi}\,Im\,(\frac{\vec{\nabla}\psi}{\psi})$$ $$i\hbar\,\frac{\partial\psi}{\partial t} = -\frac{\hbar^2}{2m}\nabla^2\psi + V(\vec{r})\,\psi$$ |

TAB.I refers to *our own approach*, whose basic equations are encoded in the structure itself of Schrödinger's time-independent equation and provide the *exact trajectories of point-particles* of assigned energy *E*, *piloted* by de Broglie matter waves of amplitude *R* (whose objective reality is testified by its diffractive properties);

TAB.II, referring to the *Bohmian approach*, provides, on the other hand, a set of *probability flow-lines* (resulting from the entire ensemble of stationary eigen-solutions), built up by the simultaneous solution of Schrödinger's time-dependent equation: *a picturesque completion of the Copenhagen interpretation.*

To our simple *Helmholtz coupling* between the exact trajectories of de Broglie's (*objective and experimentally well sound*) mono-energetic waves there corresponds the inextricable, probabilistic *entangling* among different parts of the whole system involved by Born's Wave Function and by Bohm's guidance equation.

We may conclude that our approach, allowing an *exact* Wave Mechanics running as close as possible to Classical Mechanics, is not a particular case of the *probabilistic* Bohmian Mechanics, but a *basically different vision of physical reality*.

# REFERENCES

[1]   A. Orefice, R. Giovanelli and D. Ditto, *Found. Phys.* **39**, 256 (2009)
[2]   A. Orefice, R. Giovanelli and D. Ditto, Ref.[30], Chapter 7, pag.425-453 (2012)
[3]   A. Orefice, R. Giovanelli and D. Ditto,
      *Annales de la Fondation Louis de Broglie*, **38**, 7 (2013)
[4]   A. Orefice, R. Giovanelli and D. Ditto,
      *Annales de la Fondation Louis de Broglie,* **40**, 1 (2015)
[5]   H. Goldstein, *Classical Mechanics*, Addison-Wesley (1965)
[6]   Gaussian beam - Wikipedia, the free encyclopedia
[7]   S. Nowak and A. Orefice*, Phys. Fluids B* **5**, 1945 (1993)
[8]   S. Nowak and A. Orefice*, Phys. Plasmas* **1**, 1242 (1994)
[9]   C. Honoré, P. Hannequin, A. Truc and A. Quéméneur, *Nuclear Fusion* **46**, S809 (2006)
[10]  L. de Broglie, *Compt. Rend. Acad. Sci.* **177**, pg. 517, 548 and 630 (1923)
[11]  L. de Broglie, *Annales de Physique* **3**, 22 (1925) (Doctoral Thesis of 1924)
[12]  E. Schrödinger, *Annalen der Physik* **79**, pg. 361 and 489 (1926)
[13]  E. Schrödinger, *Annalen der Physik* **81**, 109 (1926)
[14]  E. Persico, *Fundamentals of Quantum Mechanics*, Prentice-Hall, Inc. (1950)
[15]  A. Messiah, *Mécanique Quantique*, Dunod (1959)
[16]  C. J. Davisson and L. H. Germer, *Nature* **119**, 558 (1927)
[17]  A. S. Sanz and S. Miret-Artés, *A trajectory Description of Quantum Processes*,
      Springer (2012)
[18]  A. Orefice and S. Nowak, *Phys. Essays* **10**, 364 (1997)
[19]  M. Born, *Zeitschrift für Physik* **38**, 803 (1926)
[20]  N. P. Landsman, *Compendium of Quantum Physics*,
      ed. by F. Weinert, K. Hentschel, D. Greeberger and B. Falkenberg, Springer (2008)
[21]  M. Born, *Atomic Physics,* Blackie & Son Ltd (1935)
[22]  D. J. Bohm, *Phys. Rev.* **85**, 166 (1952)
[23]  D. J. Bohm, *Phys. Rev.* **85**, 180 (1952)
[24]  D. J. Bohm and B. J. Hiley, *Found. Phys.* **5**, 93 (1975)
[25]  D. J. Bohm and B. J. Hiley, *Found. Phys.* **12**, 1001 (1982)
[26]  D. J. Bohm and B. J. Hiley, *Found. Phys.* **14**, 255 (1984)
[27]  D. J. Bohm, B. J. Hiley, *Physics Reports* **172**, 93-122 (1989)
[28]  P. R. Holland, *The Quantum Theory of Motion*, Cambridge University Press (1992)
[29]  D. Dürr and S. Teufel, *Bohmian Mechanics*, Springer -Verlag (2009)
[30]  A.A.V.V., *Quantum Trajectories*, ed. by Pratim Kuman Chattaraj, CRC Press (2011)
[31]  A.A.V.V., *Quantum Trajectories*, ed. By K. H. Hughes and G. Parlant,
      CCP6, Daresbury (2011)
[32]  A.A.V.V., *Applied Bohmian Mechanics: from Nanoscale Systems to Cosmology*,
      ed. by X. Oriols and J. Mompart, Pan Stanford Publishing (2012)
[33]  A. Benseny, G. Albareda, A. S. Sanz, J. Mompart and X. Oriols*,
      Eur.Phys. J. D*, **68**:286 (2014)